\documentclass{revtex4}

\usepackage{graphicx}
\def\beq{\begin{equation}}
\def\eeq{\end{equation}}
\def\bea{\begin{eqnarray}}
\def\eea{\end{eqnarray}}

\begin{document}
\bibliographystyle{revtex}

\preprint{ }

\title{Determination of the muon Yukawa coupling\\ 
at high energy $e^+e^-$ linear colliders}

\author{Marco Battaglia}
\email[]{Marco.Battaglia@cern.ch}
\author{Albert De Roeck}
\email[]{Albert.De.Roeck@cern.ch}
\affiliation{CERN, CH-1211 Geneva 23, Switzerland}

\date{\today}

\begin{abstract}
\vspace*{0.25cm}
The perspectives for the observation of the rare decay $H^0 \rightarrow \mu^+\mu^-$ 
decay and the determination of the muon Yukawa coupling at a TeV-class and at a 
multi-TeV $e^+e^-$ linear colliders are discussed. The signal for the decay can be 
obtained at $\sqrt{s}$=0.8~TeV and a first estimate of the coupling derived. A linear 
collider operating at 3~TeV, with high luminosity, is able to improve the accuracy on 
this couplings to 4\% to 11\% for 120~GeV$<M_H<$150~GeV.
\end{abstract}

\maketitle

\section{Introduction}

Understanding the mechanism of mass generation is one of the main quests for
scientific research, today and in the coming decades. In the Standard Model (SM) 
the Higgs mechanism is held responsible for breaking the electro-weak symmetry and 
providing particles with their mass through their interactions with the Higgs 
field. Beyond the observation of a Higgs boson, the fundamental test that the 
Higgs couplings to particles scale as their masses needs to be verified experimentally,
for gauge bosons and fermions separately. The potential of a TeV-class high luminosity 
$e^+e^-$ linear collider (LC) has been already established for $W$ and $Z$ gauge 
bosons, $t$, $b$ and $c$ quarks and for the $\tau$ lepton~\cite{Battaglia:2000jb}. 
We now discuss the 
possibility of measuring the muon Yukawa coupling by the determination of the 
branching fraction of the rare Higgs decay $H^0 \rightarrow \mu^+\mu^-$. This
measurement would enable to complete the test of the coupling scaling for gauge 
bosons, quarks and leptons separately and thus ensuring that the observed Higgs boson 
is indeed responsible for the mass generation of all elementary particles. We present 
results for a TeV-class LC operating at a centre-of-mass energy $\sqrt{s}$=0.8~TeV 
and for a multi-TeV LC with $\sqrt{s}$=3.0~TeV. We considered data sets of 1~ab$^{-1}$, 
corresponding to about two years of data taking at the nominal {\sc Tesla} luminosity, 
and 5~ab$^{-1}$, corresponding to five years at the {\sc Clic} project luminosity, 
respectively.

In the SM, the branching fraction BR($H^0 \rightarrow \mu^+\mu^-$) is of the order of 
only $10^{-4}$, for Higgs masses $M_H$ below the $WW$ threshold and even lower above it. 
But the signal can be efficiently reconstructed from the 
invariant mass of two oppositely charged muons both in the Higgs-strahlung, 
$e^+e^- \rightarrow H^0Z^0$, and in the $WW$ fusion, 
$e^+e^- \rightarrow H^0 \nu_e \bar{\nu_e}$, processes and isolated from the backgrounds. 
The signal reconstruction and background rejections have been studied using the 
{\sc Simdet} parametric detector simulation program~\cite{Pohl:1999uc}, tuned to the 
expected responses for detectors at TeV and multi-TeV LCs. In particular, a muon 
momentum resolution $\delta p/p^2=5\times10^{-5}$~GeV$^{-1}$ has been assumed. The 
signal and background cross sections have been obtained using the {\sc CompHEP} 
program~\cite{Pukhov:1999gg}, while the SM BR($H^0 \rightarrow \mu^+\mu^-$) has been 
computed with {\sc HDecay}~\cite{Djouadi:1998yw}. The signal events have been generated 
using the {\sc Pythia-6} Monte Carlo~\cite{Sjostrand:2001yu} while parton level 
generation of the background processes has been performed with {\sc CompHEP} and the 
events have then been hadronised using {\sc Pythia}~\cite{Belyaev:2000wn}. 

\section{$H^0 \rightarrow \mu^+\mu^-$ at a TeV-class LC}

While most of the profile of a light Higgs boson is best studied with lower energy 
operation of a LC, 300~GeV$<\sqrt{s}<$500~GeV, there are advantages in performing some 
measurements at energies around 1~TeV. The total Higgs production cross section is 
within a factor of two from that at the peak for the Higgs-strahlung process and this 
is more than compensated by the increase in the achievable luminosity at these larger 
energies. Further the $ZZ^*$ background is significantly decreased. We have considered
$\sqrt{s}$=0.8~TeV with an integrated luminosity ${\cal{L}}$=1~ab$^{-1}$, corresponding 
to 1.75$\times 10^5$ Higgs bosons produced in the $e^+e^- \rightarrow H \nu \bar \nu$ 
channel and about 40 $H \rightarrow \mu^+\mu^-$ decays, for $M_H$=120~GeV.

\begin{figure}[t]
\begin{tabular}{l r}
\includegraphics[width=0.51\textwidth,height=0.35\textwidth]{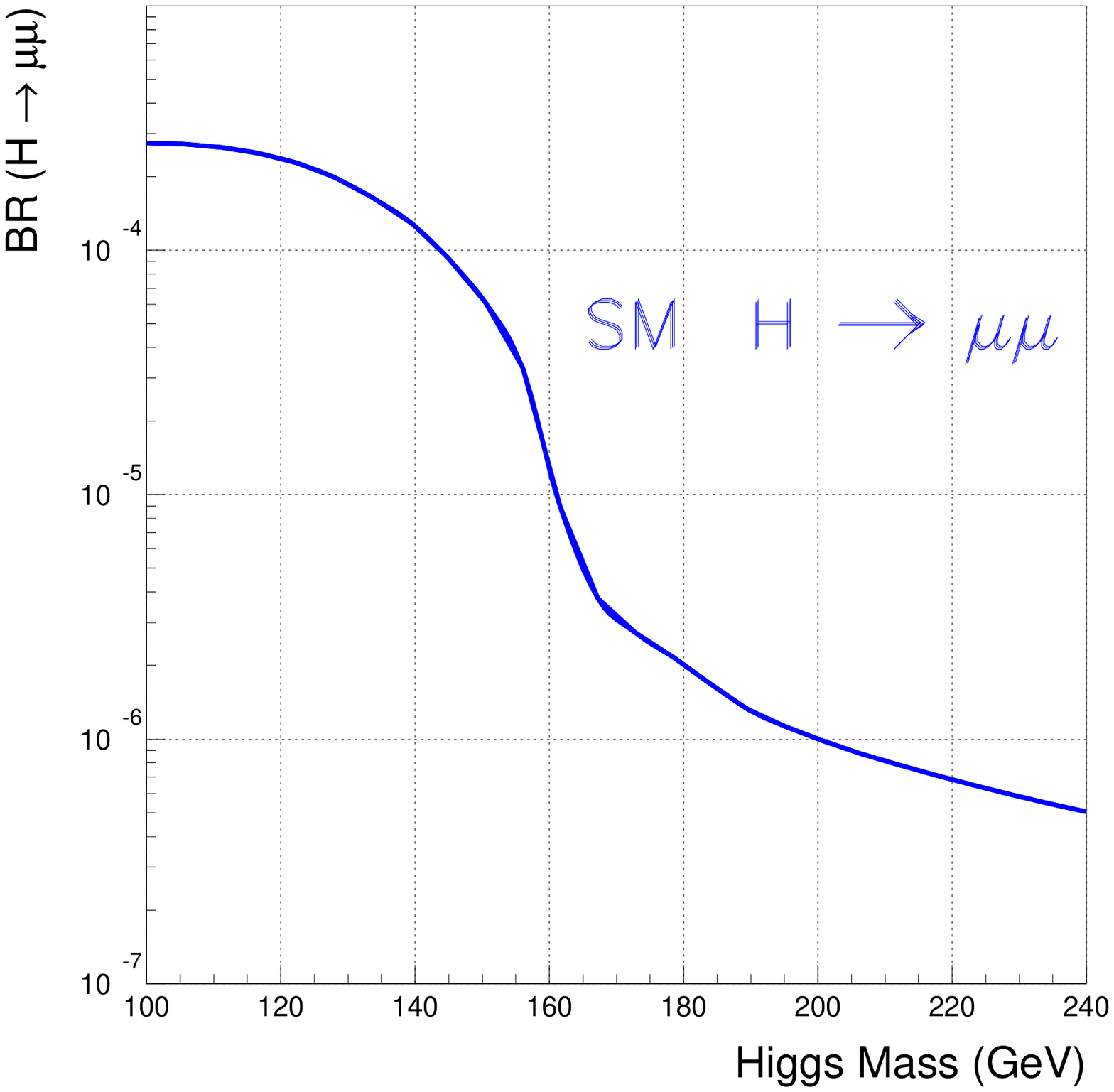} &
\includegraphics[width=0.51\textwidth,height=0.35\textwidth]{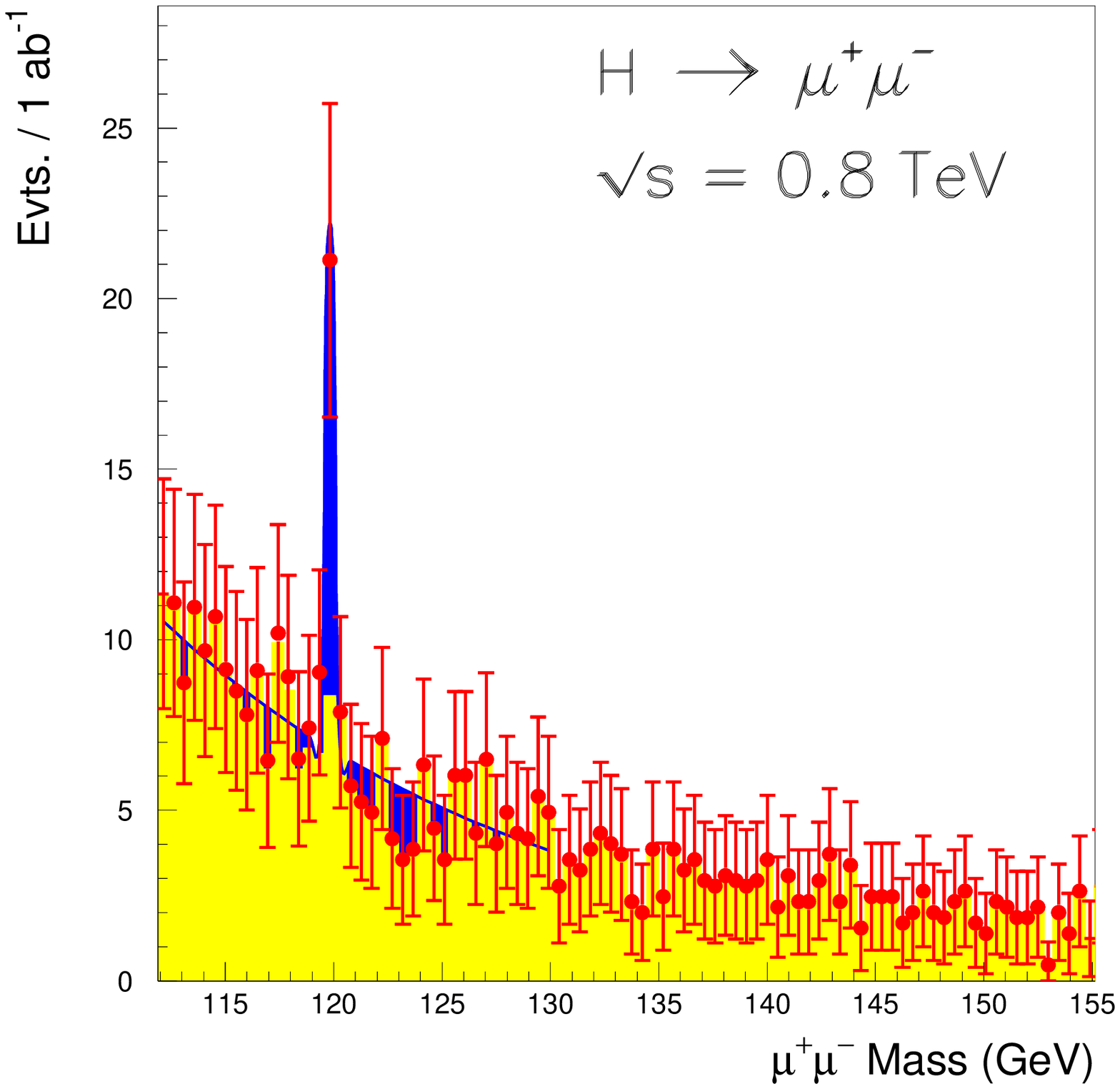} \\
\end{tabular}
\caption{Left: SM prediction for BR($H \rightarrow \mu^+\mu^-$) as a function of the 
Higgs boson mass. Right: 
distribution of the di-muon invariant mass $M_{\mu\mu}$ for $M_H$=120~GeV. 
The points with error bars represent 1~ab$^{-1}$ of data at $\sqrt{s}$=0.8~TeV 
and the continuous line the result of the fit used to extract the number of signal 
events.}
\label{fig:fig1}
\end{figure}
  
The analysis starts from events with two oppositely charged, identified muons and 
significant missing energy and assumes an accurate knowledge of the Higgs mass, obtained
at the {\sc Lhc} and lower energy LC operation. 
The main background from $WW \rightarrow \mu \nu \mu \nu$ 
is reduced by cuts on the di-muon recoil mass $M_{\mu\mu}^{recoil}$ and energy 
$E_{\mu\mu}$.
The total background has been estimated including also the $ZZ\nu\bar\nu$, 
$WW\nu\bar\nu$ and the inclusive $\mu\mu\nu\bar\nu$ processes, evaluated without the 
Higgs contribution. 
The resulting di-muon invariant mass is shown in Figure~\ref{fig:fig1}.
The $H \rightarrow \mu\mu$ signal can be extracted from the underlying background 
with more than 5$\sigma$ significance, for $M_H \simeq$~120~GeV. From the event rate, 
the product of production cross section and $\mu\mu$ decay branching fraction 
$\sigma(e^+e^- \rightarrow H\nu\nu)\times{\mathrm{BR}}(H \rightarrow \mu\mu)$ 
is measured with $\simeq 30$~\% accuracy.

\section{$H^0 \rightarrow \mu^+\mu^-$ at a multi-TeV LC}

Multi-TeV $e^+e^-$ collisions provide a large sample of Higgs bosons, produced in 
the $WW$ fusion process, becoming the dominant Higgs production mode since 
$\sigma_{H\nu\bar\nu} \propto log~\frac{s}{M^2_H}$. However, beyond $\sim$~3~TeV the 
further gain in production cross section is largely lost due to the reduced acceptance, 
due to the forward Higgs boson production. A data set of 5~ab$^{-1}$ at 
$\sqrt{s}$=3~TeV corresponds to a sample of 2.7$\times 10^6$ Higgs bosons and 
650 $H \rightarrow \mu\mu$ decays, for $M_H$=120~GeV.
\begin{figure}[t]
\begin{tabular}{c c}
\includegraphics[width=0.51\textwidth,height=0.35\textwidth]{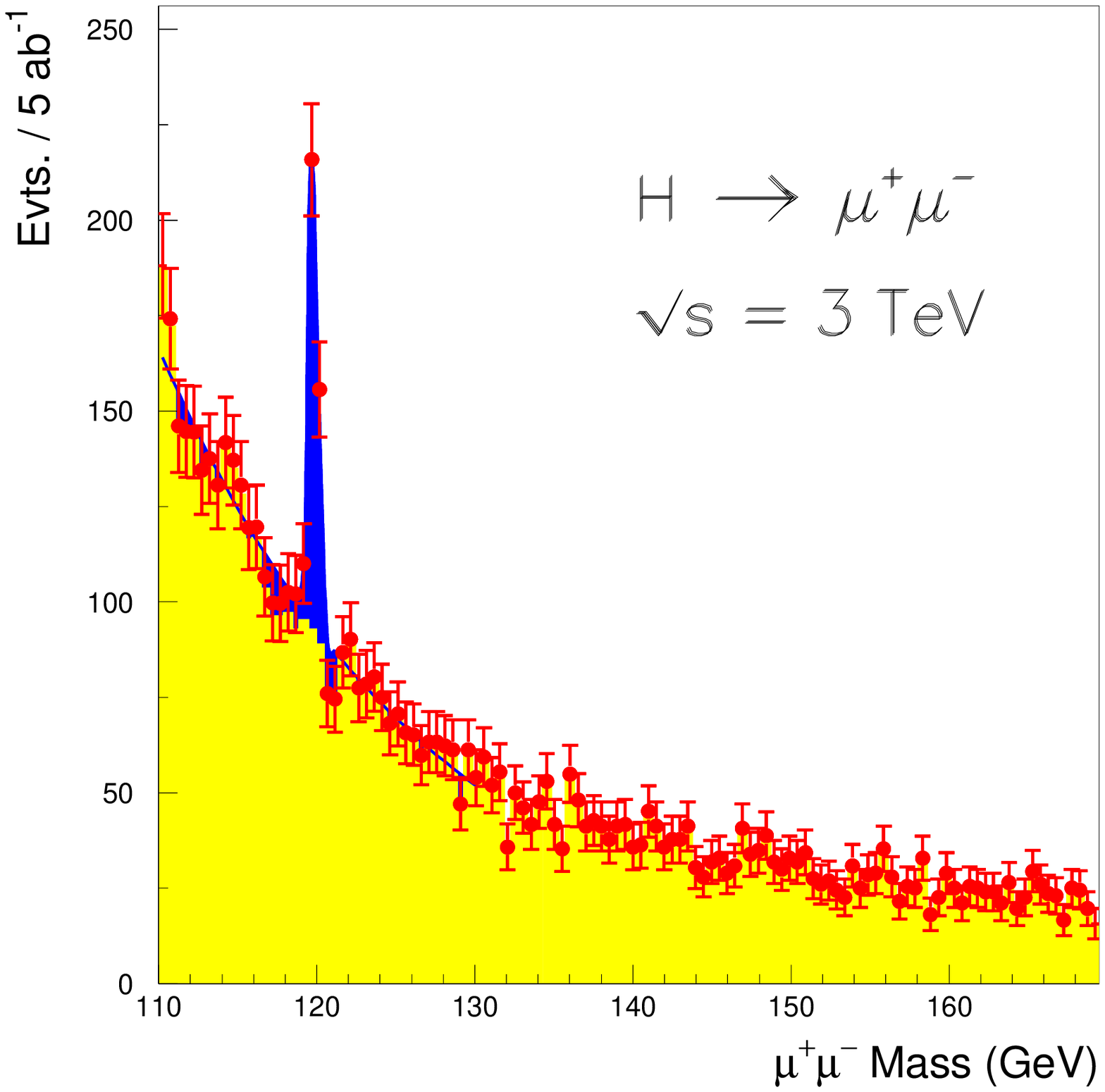}&
\includegraphics[width=0.51\textwidth,height=0.35\textwidth]{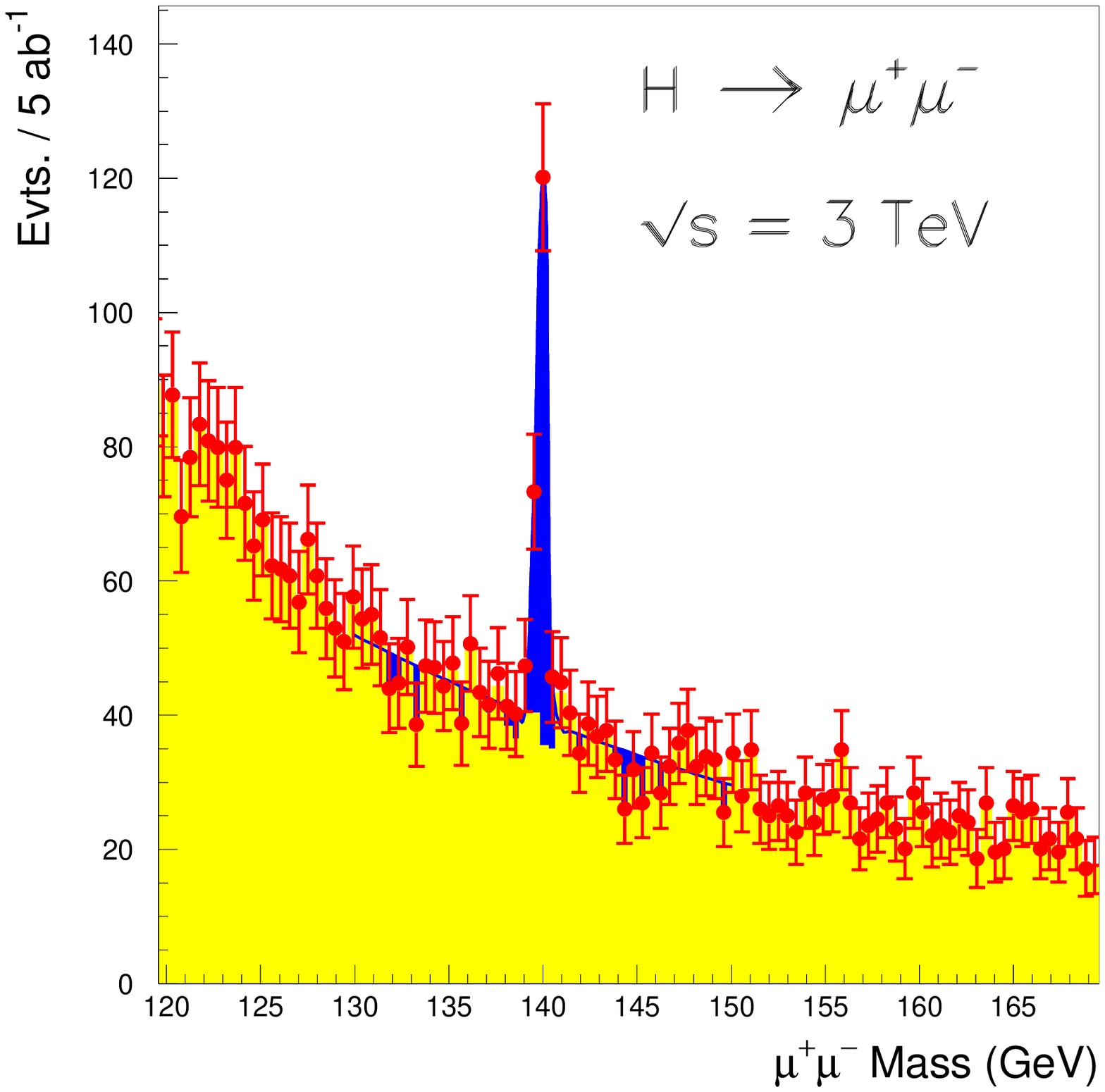}\\
\end{tabular}
\caption{Distribution of the di-muon invariant mass $M_{\mu\mu}$ for two data sets 
with $M_H$=120~GeV (left) and $M_H$=140~GeV (right). The points with error bars 
represent 5~ab$^{-1}$ of data at $\sqrt{s}$=3~TeV and the continuous lines the 
results of the fits used to extract the numbers of signal events.}
\label{fig:fig2}
\end{figure}

The analysis is again based on events with two identified muons and the background is 
suppressed in part by selection cuts on the di-muon recoil mass and energy. The 
$H \rightarrow \mu \mu$ signal is clearly visible in the di-muon invariant mass 
distribution above the background. A signal significance of more than $5\sigma$ 
is obtained up to $M_H \simeq$ 155~GeV. The number of signal events is extracted from a 
fit to the di-muon invariant mass where the signal is modelled by a Gaussian distribution
peaked at the nominal Higgs mass and the background by a polynomial curve fitted on the 
peak side bands (see Figure~\ref{fig:fig2}). 
The accuracies on the product of production cross section and $\mu\mu$ decay branching 
fraction  derived from the fitted number of signal events, are summarised in 
Table~\ref{tab:tab1} for different values of $M_H$.
\begin{table}[t]
\caption{Statistical accuracy on the product 
$\sigma(e^+e^- \rightarrow H\nu\bar\nu)\times{\mathrm{BR}}(H \rightarrow \mu\mu)$ and 
the $g_{H\mu\mu}$ coupling expected with an integrated luminosity of 5~ab$^{-1}$ at 
$\sqrt{s}$=3~TeV, for different values of the Higgs boson mass $M_H$.}
\begin{tabular}{|l|c|c|c|}
\hline
$M_H$ & 120~GeV & 140~GeV & 150~GeV \\ \hline \hline
$\delta$($\sigma_{H\nu\bar\nu}\times$BR)/($\sigma_{H\nu\bar\nu}\times$BR) & 
0.072 & 0.120 & 0.210  \\ \hline
$\delta g_{H\mu\mu}/g_{H\mu\mu}$ & 0.040 & 0.062 & 0.106  \\ \hline
\end{tabular}
\label{tab:tab1}
\end{table}

\section{Discussion and Conclusions}

The $g_{H\mu\mu}$ coupling can be extracted from the measured branching fractions with 
an independent measurement of the production cross section, obtained using the dominant
$b \bar b$ and/or $WW^*$ Higgs boson decay modes. An uncertainty of 2\% is assumed 
for the production cross section, extrapolating results from lower energies. The 
resulting accuracies on the muon Yukawa coupling are given in Table~\ref{tab:tab1}.

A muon collider (FMC), running at energies around the $H$ mass, can determine the product
$\sigma(\mu\mu \rightarrow H) \times {\mathrm{BR}}(H \rightarrow b \bar b)$ to a 
statistical accuracy of 3.5\% to 0.2\% for $M_H \simeq$~120~GeV, depending on the 
assumed luminosity and beam energy spread~\cite{Gunion:1997sk,Autin:1999ci}.  By taking 
the ${\mathrm{BR}}(H \rightarrow b \bar b)$ measured at the LC to 2.5\%, the 
combination of the 
FMC+NLC data allows to extract the $g_{H\mu\mu}$ coupling with an accuracy of
$\simeq 2.2-1.3\%$. To this accuracy, the $b \bar b$ branching fraction 
contributes with comparable or dominant uncertainty depending on the FMC luminosity 
assumption. 

The decay $H \rightarrow \mu^+\mu^-$ may also be investigated at hadron colliders. 
However, a first study~\cite{Plehn:2001qg} has shown that the {\sc Lhc} could get a 
$5\sigma$ signal of this decay only with an integrated luminosity in excess of 
2~ab$^{-1}$. Therefore, it may become relevant only for a high luminosity upgrade. 
At higher energies, a {\sc VLhc} would be able to get a $5\sigma$ evidence with about 
0.25~ab$^{-1}$ and a $g_{H\mu\mu}$ coupling determination with 10-14\% accuracy for 
120~GeV$<M_H<$140~GeV at $\sqrt{s}$=200~TeV.

The measurements of the muon Yukawa coupling allow to extend the test of the Higgs 
mechanism of mass generation to the lepton sector by verifying that 
$\frac{g_{H\mu\mu}}{g_{H\tau\tau}} = \frac{M_{\mu}}{M_{\tau}}$. The $g_{H\tau\tau}$ 
coupling can be determined at the LC with an accuracy of 3\% to 5\% for 
120~GeV$<M_H<$140~GeV, already at $\sqrt{s}$=300~GeV-500~GeV~\cite{Battaglia:2000jb}. 
The accuracies of the determination of the two couplings being comparable, the test of 
the mass scaling can be obtained with a precision of 5.0\% to 8.0\% for 
120~GeV$<M_H<$140~GeV for the case of a LC and of 3.3\% to 3.7\% for $M_H$=120~GeV at a 
FMC, depending on the luminosity.

In summary, we discussed the measurement of the muon Yukawa coupling by the 
determination of the branching fraction of the rare Higgs decay 
$H^0 \rightarrow \mu^+\mu^-$ at $\sqrt{s}$ = 0.8~TeV and 3.0~TeV. 
The $H \rightarrow \mu^+ \mu^-$ signal can be observed at the lower energy for $M_H$ = 
120~GeV and the $g_{H\mu\mu}$ coupling measured to $\simeq$15\%. This accuracy can be 
to pushed to 4-11\% for the wider Higgs mass range 120~GeV~$< M_H <$~150~GeV with 
$e^+e^-$ collisions at $\sqrt{s}$=3~TeV.
The anticipated accuracy of a multi-TeV LC in the determination of $g_{H\mu\mu}$ 
is comparable to that achievable at a muon collider operating at the Higgs mass 
peak and more than two times better than at a 200~TeV VLHC hadron collider. 

%

\bibliography{e3066}

\end{document}